\documentstyle[12pt]{article}

\def\'#1{{\accent19\ifx #1i \i\else #1\fi}}

\def\be{\begin{equation}}
\def\ee{\end{equation}}
\def\bea{\begin{eqnarray}}
\def\eea{\end{eqnarray}}
\newcommand{\boldmathalpha}{\mbox{\boldmath$\alpha$\unboldmath}}

\newcommand{\boldmathsigma}{\mbox{\boldmath$\sigma$\unboldmath}}

\newcommand{\boldmathpi}{\mbox{\boldmath$\pi$\unboldmath}}

\newcommand{\boldmathnabla}{\mbox{\boldmath$\nabla$\unboldmath}}

\newbox\Ancha
\catcode`@=11
\newdimen\ex@
\title{ Compactification with $U(1)$ magnetic field within Dirac supersymmetry}
\author{J. Besprosvany and M. Moreno}
\date{Instituto de F\'{\i}sica, Universidad Nacional Aut\'onoma de M\'exico,
Apartado Postal 20-364, M\'exico 01000, D. F., M\'exico}
\begin{document}
\maketitle
%\vfill\eject
%\vspace{1in}
%\vskip30pt
%leftline{Short title:}
%\vskip40pt
%\centerline{PACS: 02.20.Qs, 03.65.Fd, 03.65.Pm}
%\vfill\eject
%\vfill\eject
\jot = 1.5ex
\def\baselinestretch{1.10}
\parskip 5pt plus 1pt
\begin{abstract}
%137 palabras!
%04.50.+h Gravity in more than four dimensions, Kaluza-Klein
%theory, unified field theories; alternative theories of gravity
%(see also 11.25.Mj Compactification and four-dimensional models)

%11.10.Kk Field theories in dimensions other than four (see also

%98.80.Cq Particle-theory and field-theory models of the early
%Universe (including cosmic pancakes, cosmic strings, chaotic
%phenomena, inflationary universe, etc.)

%12.60.-i Models beyond the standard model see also 12.10 Unified
%field theories and models

%11.25.Mj Compactification and four-dimensional models

We consider     Dirac-supersymmetric  interactions, which produce
CP-conserving separation of positive and negative energy solutions
in the Dirac equation in order to investigate an alternative to the
 Kaluza-Klein mechanism. We review conditions under which
separation is possible into free particle and compactified
behaviors in different dimensions,
 with attention to  spin degrees of freedom. We show a $U(1)$
constant magnetic field produces such kind of behavior; an
explicit treatment is given to the  6-$d$ to 4-$d$
 and   4-$d$ to 2-$d$ breaking cases and  the spectrum is obtained. A dynamical mass-creation
 mechanism is suggested from the procedure.
%is shown to predict the  $SU(3)\times SU(2)_L\times U(1)$ symmetry and a three
%flavor symmetry.
\end{abstract}
%\vskip 1cm \centerline{PACS: 03.65.Bz, 03.70.+k, 12.60.-i}
%03.65.Ca Formalism
%03.65.Bz Foundations, theory of measurement, miscellaneous theories (including Aharonov-Bohm effect, Bell inequalities,
%12.10.Dm Unified theories and models of strong and electroweak
%03.70.+k Theory of quantized fields (see also 11.10 Field theory)
%12.60.-i Models beyond the standard model
 \vskip 1cm
 \begin{center}
% VERSION PRELIMINAR
 \end{center}
 \baselineskip 22pt \vfil\eject
\section{Introduction}
The Kaluza-Klein idea has been  a  promising and useful assumption
in modern physical research  whose aim is the unification of
forces.
 Although no experimental indication exists to presuppose its
 existence,
it remains a useful working hypothesis in various theories.
 The original Kaluza-Klein idea
proposes a fifth dimension that accounts for the electromagnetic
interaction in the framework of an extended general relativity,
and thus has succeeded in presenting gravity and electromagnetism
in a unified picture. Additional dimensions have been proposed to
account for the other fundamental interactions. Indeed,  this is
the underlying aim  in applications  in  supergravity and string
theory, where additional compactified dimensions of space have
been linked to gauge interactions. In fact, arguments based on the
supersymmetry breaking scale\cite{anto} and recent developments in
string theory\cite{Dim} have opened up the possibility that
unification occurs at the electroweak scale, with the implication
that the additional dimensions might be detected through its
effect on gravity at millimeter scales.

Within an ampler view, the question why there are four and not
any other number of physical dimensions remains unanswered. To
investigate this question,  one can check whether it is possible
to construct a model in which additional dimensions lead
consistently to the same four-dimensional physics. %NEW

In particular,  the idea of higher dimensions should provide also
for mechanisms in which an assumed larger dimensional universe
transforms into the present one, with an expected explanation of
the fate of other dimensions.  The most popular related
assumption for this is Klein's idea of compactification, which
assumes dimensions become unobservable by closing on themselves
at a small length.
%applications have not yet developed to the point that its tenets
%can be subjected to experimental test
 However, the origin, dynamics,  and    range of applicability
of this process have  not been thoroughly  pursued and clarified,
as will be implied from the present work. Indeed, when
considering this mechanism, its existence has generally been taken
for granted but not the causes leading to it.

In addition,  several   methods of dimensional reduction and
compactification (see Ref.  \cite{Bai} for a review) are known but
few account successfully for the particles' spectrum,
representations and, in particular, for the demand that chiral
fermions be obtained to reproduce the quantum numbers of physical
particles\cite{Wit}.

Even when acceptable compactifications are found, there are
multiple  choices with similar four-dimensional physics,  which
points at the need of additional feasible restrictions to reduce
the number of possibilities and increase the predictability.
Gauge fields acting on particles is such a physical mechanism,
and needs to be examined. The choice of these fields is motivated
by the already known four-dimensional ones. The self-consistency
of these fields should be checked in the next stage of the study
of the problem. %NEW

 To investigate a possible
dynamical process  that generates compactification  is the main
object of this work. We will explore the idea that this
compactification is generated by the gauge fields themselves,
which we think more economical. It is natural to start by
considering the Dirac equation, which describes  basic spin-1/2
fields. Thus, the  mechanism proposed should allow for a
description of the fermion fields in the presence of usual and
compactifying interactions. We require that the latter should not
affect the workings of the first and the Dirac fields.  It is
also expected that they conserve the symmetry in the weight of
positive and negative energies so that when these are turned on
the vacuum should not be altered asymmetrically. It is
 known that such interactions exist\cite{Mar} and that they
satisfy a restriction related to the presence of the Dirac
supersymmetry\cite{DiracSuper}.

In this paper we  investigate Dirac-supersymmetric interactions
which effectively lead  to    separation into physical and
compactified dimensions and we find indeed
 a simple instance in which this is possible. This material is organized as follows:
 In Section II we review Dirac supersymmetry which
defines some of the conditions for interactions in which a
dynamical dimension separation mechanism can be possible. In
Section III we investigate separation of the Dirac-supersymmetric
equation using as example the 3+1 dimensional case. In Section IV
we consider a dimension separation mechanism in which a $U(1)$
magnetic field is assumed and which produces an effective
compactification in some dimensions. This model is considered in
both a 4-$d$ to 2-$d$ (1+1) and a 6-$d$ to 4-$d$ (3+1)
transitions. In Section V we draw some conclusions from this work.
\section{Interactions in Dirac supersymmetry}
The basic feature defining   Dirac supersymmetric interactions  is
the possibility of applying a generalized Foldy-Wouthuysen
transformation which brings the usual Dirac Hamiltonian to a form
in which positive and negative solutions  separate with equal
weight.
Explicitly, we assume the initial Dirac Hamiltonian $H$ can be
written in the form
\begin{eqnarray}
\label{DiracSup} H=Q+Q^\dagger+\lambda,
\end{eqnarray}
where $\lambda$ is a Hermitian operator and $Q$ is a fermionic
operator such that it and its adjoint $Q^\dagger$ satisfy
${Q}^2=0$ and ${Q^\dagger}^2=0$, and  also the anticommutation
relations
\begin{eqnarray}
\label{anticom}
 \{ Q,\lambda \} =0\ \ \ \ \ \{ Q^\dagger,\lambda \}=0 .
\end{eqnarray}
After applying the Foldy-Wouthuysen transformation (FWT) $H$
becomes\cite{Mar}
\begin{eqnarray}
\label{FWtrans}
 H^\prime= \frac{\lambda}{(\lambda^2)^{1/2} }
 {(\{Q,Q^\dagger \} +\lambda^2)}^{1/2}.
\end{eqnarray}
An extensive analysis of the interactions that take the form in
Eq. (\ref{DiracSup}) is given in Ref. \cite{Mar}. The resulting
ones are given in the Hamiltonian (in 4+1 dimensions)
\begin{eqnarray}
\label{HamDS}
 H={ \boldmathalpha}\cdot{ \boldmathpi}+{ \alpha}_5\pi_5+\mu\beta ,
\end{eqnarray}
where
\begin{eqnarray}
\label{pi} { \pi}_I={  p}_I+{\mathsf A}_I({\bf x})+i \beta
{\mathsf E}_I({\bf x}) \ \ \ \ I=1,2,3,5,
\end{eqnarray}
${ \alpha}_i$ are Dirac matrices with ${
\alpha}_i=\gamma_0\gamma^i$, ${ \alpha}_5=i\beta\gamma_5$,
${\mathsf A}_I({\bf x})$ and ${\mathsf E}_I({\bf x})$ are external
arbitrary fields, and $\mu$ is the mass constant, which
multiplies the corresponding term.
%se usa mathbf
\section{Separation of Dirac-supersymmetric equation}
 Our main concern
here is to study   dynamical compactification as a  dimensional
reduction effect and, in particular, the form of those
interactions which not only satisfy Dirac-supersymmetric
conditions but also allow for such mechanism. Specifically, we
need a mechanism in which
%withoutat this point asking the mechanism behind,
an interaction would confine particles in some dimensions and keep
free-particle behavior in the unconfined (physical) dimensions.
For this purpose we use as illustrative model the $3+1\rightarrow
1+1$ reduction case, studying separation of variables for $H$ in
Eq. (\ref{HamDS}). This $H$ can  be written in the form of Eq.
(\ref{DiracSup}) with the association
\begin{eqnarray}
\label{HamDSMa}
\lambda &=&\mu \beta \\
 Q &=& \left (
  \begin{array}{cc}
  0 & 0 \\
  M & 0
\end{array}
 \right ) \\
 \label{HamDSMaL}
 Q^\dagger &=& \left (
  \begin{array}{cc}
  0 & M^\dagger\\
  0 & 0
\end{array}
 \right ),
\end{eqnarray}
where  \begin{eqnarray} \label{Mterm} M=\boldmathsigma\cdot({\bf
p}+{\bf C})-i C_5,
\end{eqnarray} $C_I=A_I-i E_I$, $I=1,2,3,5$ and we use here,  unless otherwise stated, the
standard representation for the Dirac matrices\cite{Bjor}. The
expression for the Hamiltonian in Eq. (\ref{HamDS}), after
application of the FWT and after using Eqs.
(\ref{HamDSMa})-(\ref{HamDSMaL}), is
\begin{eqnarray}
\label{HamFW}
 H^\prime &=& \beta\left [\left (
  \begin{array}{cc}
  MM^\dagger+\mu^2 & 0 \\
  0 & M^\dagger M+\mu^2
\end{array}
 \right )\right ]^{1/2},
\end{eqnarray}
where the equation to solve is
\begin{eqnarray}
\label{HamFWeq}
 H^\prime \Psi =E \Psi.
\end{eqnarray}
The equality between positive and negative spectrum, except for a
sign, follows from the equal eigenvalues expected for both terms.

The square-operator  form  of the upper and lower terms of
$H^\prime$ on Eq. (\ref{HamFW}) suggests a simplification by
considering the action of  single operators. Clearly, this is
permitted when $M$ is hermitian. It is then is possible to
reformulate   Eq. (\ref{HamFWeq})
 in terms of another
  eigenvalue equation as alternative
sufficient condition, linear in $M$, and   given by
\begin{eqnarray}
\label{Eigenfacil} M \psi=E_c\psi,
\end{eqnarray}
where $\psi$ is the positive energy spinor  component of $\Psi$
since, as can be proved, both the upper and lower sides in
$H^\prime$ in Eq. (\ref{HamFWeq})  have the same eigenvalues.  In
fact, this equation is valid also in the case $M$ is not
hermitian and therefore $E_c$ complex, for  we only require
$|E_c|^2 = E^2-\mu^2$.

The linear form of  Eq. (\ref{Eigenfacil}) hints at a possible
separation of  degrees of freedom corresponding to different
dimensions. The separation of this equation (in the stationary
case) into independent components describing different dimensions
$M=M_1+M_2$ is limited, for
 the  form
of $M$ implies that even if use, say, in $H_1$ potentials with
restrained dependence on the coordinates we generally obtain
$[M_1,M_2]\neq 0$. %As illustrative model, we first study the $3+1
%\rightarrow 1+1$ separation.
 Still, we can find a
partial separation of the 3+1 Eq. \ref{Eigenfacil} into  1+1
dimensional terms and  two additional dimensional terms (1+1
scalar). We choose the 1+1 spatial coordinate as $z$, with $
x_1=x$, $x_2=y$, $x_3=z$, and
\begin{eqnarray}
 \label{M1}
M_1=\sigma_z [p_z+C_z(z)],
\end{eqnarray}
\begin{eqnarray}
 \label{M2}
M_2=\boldmathsigma_b \cdot [{\bf p}_b+{\bf C}_b(x,y)],
\end{eqnarray}
 where $b$ represents
the $(\hat x, \hat y)$ directions, and ${\bf
C}_b(x,y)=(C_x(x,y),C_y(x,y))$, with dependences of ${\bf C}_b$
and $C_z$  as specified.   The Anzats
\begin{eqnarray}
\label{anzats}
 \psi =\left ( \begin{array} {c} g_1(x,y) f_1(z)\\ g_2(x,y) f_2(z) \end{array}\right )
\end{eqnarray}
leads to the partial separation of Eq. \ref{HamFWeq}
\begin{eqnarray}
\label{sepa1}
 M_1\left ( \begin{array} {c} f_1(z)\\ f_2(z) \end{array}\right
 )=
 \left ( \begin{array} {c} E_1 f_1(z)\\  {E_1}^\prime f_2(z) \end{array}\right
 ),
\end{eqnarray}
where we have used the diagonal character (in spin space) of the
$M_1$ component which allows for cancellation of the $g_i(x,y)$
in Eq. (\ref{sepa1}). We further assume that the upper and lower
components share the same solutions, so that $f_1(z)=f_2(z)$.
Then, $E_1=-{E_1}^\prime$.  We note solutions can be interpreted
as positive and negative chirality components in 1+1 space. The
other part of  Eq. \ref{HamFWeq} has consequently the form
\begin{eqnarray}
\label{sepa2}
 M_2\left ( \begin{array} {c} g_1(x,y)\\ g_2(x,y) \end{array}\right
 )=\left ( \begin{array} {c} (E_c-E_1)g_1(x,y)\\  (E_c+E_1) g_2(x,y) \end{array}\right )
\end{eqnarray}
 and the $z$ dependence can be divided out. The $M_2$  part plays the role
of a scalar interaction in  $1+1$ space. The absence of the
$f_i(z)$ functions implies the eigenvalues $E_c$ depend on the
$f_i$ solutions only through $E_1$. Thus, this separation is only
partial yet  sufficient for our purposes because it decouples at
least one group of dimensions. It is clear that the separation
depends on the presence of a diagonalizable component  as $M_1$,
and on the  specific  spatial dependence of the potentials in
$M_1$, $M_2$, in accordance to the Lorentz index of the
$\alpha_i$ matrices. In this case, it is the need to account for
the   spin degree of freedom that  requires the additional
condition that $M_1$ be diagonalizable. We also find that in
passing from the higher dimension to the lower, the original spin
is reinterpreted  and forms the  chiral components in the lower
dimension.
\section{Dirac supersymmetry in higher dimensions}
A thorough analysis of the process of Dirac supersymmetric
Hamiltonians breaking into lower dimensional components requires
an understanding of a Dirac-matrices construction which exhibits
Dirac supersymmetry. This analysis is performed in detail in Ref.
\cite{MatRos} and here we reproduce some useful results. In
general, a Clifford algebra is defined through the anticommutation
relations
\begin{eqnarray}
\label{anticomm} \alpha_i \alpha_j+\alpha_j\alpha_i=2\delta_{ij},
\end{eqnarray}
where $i,j=0,1,...,d-1$ and $d$ is an integer. The main relation
that we will use is a  recursive formula that provides the
elements of the even-$d$ algebra in terms of the $d-2$ algebra.
This is
\begin{eqnarray}
\label{recualg}
 \alpha_0 & = \sigma_3 \otimes \hat 1=& \left (
  \begin{array}{cc}
  \hat 1 & 0 \\
  0 & -\hat 1
\end{array} \right )\\
\alpha_i & =  \sigma_1 \otimes \hat \alpha_i=& \left (
  \begin{array}{cc}
  0 & \hat\alpha_i \\
 \hat\alpha_i & 0
\end{array}\right )\ \ \ \ \ \ \ \ \ 1\leq i \leq d-2 \\
 \alpha_{d-1} & =  \sigma_1 \otimes \hat 1=& \left (
  \begin{array}{cc}
  0 & \hat\alpha_0 \\ \label{recualgend}
  \hat\alpha_0 & 0
\end{array}\right )\\ \label{recualgF}
\alpha_{d} & =  \sigma_2 \otimes \hat 1=& \left (
  \begin{array}{cc}
  0 &  -i\hat 1 \\
  i\hat 1 & 0
\end{array}\right ),
 \end{eqnarray}
 where the caret denotes the lower-dimensional algebra. The latter
element  $\alpha_d$  is obtained from the definition
\begin{eqnarray}
\label{alphad}
 \alpha_d=e^{-i(d/2)\pi/2}\alpha_0...\alpha_{d-1},
\end{eqnarray}
and it extends the algebra from even $d$  to odd $d+1$.

The
 Lorentz generator antisymmetric
tensors, generalized to any dimension, can be deduced from Eqs.
(\ref{recualg})-(\ref{recualgF}) and the definition of the
$\gamma$ matrices $\gamma^0=\beta$, $\gamma^i=\beta\alpha_i$.
They can be shown to be
\begin{eqnarray}
\label{antismunu}
 \sigma^{\mu\nu}=\frac{i}{2}[\gamma^\mu,\gamma^\nu]= \left (
  \begin{array}{ccc}
  0& i \sigma_1 \otimes \hat \alpha_i & i \sigma_2 \otimes \hat 1  \\
   & 1 \otimes \hat \sigma^{ij} &   \sigma_3 \otimes \hat \alpha_i\\
     & &  0
\end{array}
 \right ) \begin{array}{c} 0\\1\leq i \leq d-1\\d, \end{array}
\end{eqnarray}
where the rows' labels are given and the column labels follow the
same order (the elements below the diagonal are minus the
transpose of those above).

 In general, by investigating the general structure of all
supersymmetric terms, one can also show all Dirac-supersymmetric
interactions have the form
\begin{eqnarray}
\label{genform}
Q&=&\frac{1}{2}(\sigma_1+i\sigma_2)\otimes \hat q \\
Q^\dagger&=&\frac{1}{2}(\sigma_1-i\sigma_2)\otimes \hat q^\dagger,
\end{eqnarray}
where $\hat q$ represents any  interaction   in the $d-2$ space.
With this characterization one may proceed  from the
transformation that departs from Eq. (\ref{DiracSup}), goes
through $H^\prime$ in Eq. (\ref{FWtrans}) and leads  to Eq.
(\ref{HamFW}), an expression containing of $Q+Q^\dagger$ and
$\lambda$, and use the preceding section to select interactions
that allow for dimensional separation.

\section{$U(1)$ magnetic field as Dirac-supersymmetric interaction}
A magnetic field derived from a general $U(1)$ interaction
satisfies both conditions of Dirac supersymmetry and
separability. It is possible to describe it in  any dimension in
terms of an expression for $\boldmathpi$ of the form of  Eq.
(\ref{pi}).

\subsection{$3+1 \rightarrow   1+1$ reduction}
In  the $3+1$ dimensional case
 a constant magnetic field along the $\hat z$ direction
\begin{eqnarray}
\label{3dmag}
{\bf A} =-\frac{1}{2} {\bf  r} \times {\bf  B}
\end{eqnarray}
corresponds to the real elements in Eq. (\ref{HamDS})  ${\mathsf
A}_I=e A_I$, $I=1,2,3,$ where $e$ is the $U(1)$ coupling constant
(corresponding to the  electric charge) and the ${\bf A}$
components are
\begin{eqnarray}
\label{A3d}
A_1=A_x=-\frac{1}{2} y B,\ \ \ A_2=A_y=\frac{1}{2} x B, \ \ \ A_3=A_z=0.
\end{eqnarray}

The Hamiltonian $H^\prime $ (Eq. (\ref{FWtrans}))
 results in
\begin{eqnarray}
\label{KKmag}
H^\prime= \alpha_0
 {[ (\boldmathalpha\cdot\boldmathpi)^2+\mu^2 ]}^{1/2} .
\end{eqnarray}
That the magnetic field interaction chosen in Eq. (\ref{A3d})
 separates  in the sense of Eqs. (\ref{sepa1}) and
(\ref{sepa2}) can be seen from the  construction of $H^\prime$ in
Eq. (\ref{KKmag}). $H^\prime$ is composed from a term multiplying
the unit matrix and the term
\begin{eqnarray} \label{KKmagpp}
   (\boldmathalpha\cdot\boldmathpi)^2 &=& \alpha_i \alpha_j \pi_i \pi_j \\
                                      &=& (\frac{1}{2} \{ \alpha_i,\alpha_j \} +
 \frac{1}{2}[\alpha_i,\alpha_j]) \pi_i \pi_j   \\
\label{KKmagppL} &=&  \boldmathpi\cdot\boldmathpi+ i\sigma_{ij}
\pi_i \pi_j,
 \end{eqnarray}
where use has been made of Eqs. (\ref{anticomm}) and
(\ref{antismunu}). The term
\begin{eqnarray}
\label{sepaKK}i\sigma_{ij} \pi_i \pi_j=i 1\otimes \hat\sigma_{ij}\pi_i \pi_j,
\end{eqnarray}
where Eq. (\ref{antismunu}) has been used, gives rise to a
separable equation, which will be shown explicitly in the
Appendix.

%It follows that it is
%possible to directly extract each of the positive-negative energy
%components which in the $ 3+1\rightarrow 1+1$. *optional
 Squaring of $H^\prime$ of  Eq. (\ref{KKmag}) in Eq. (\ref{HamFWeq}) and further reduction
lead to the equation
\begin{eqnarray}
\label{equMAg3} ({\bf p}\cdot {\bf p} -2ie {\bf A}\cdot {
\boldmathnabla} + e^2 {\bf A} \cdot {\bf A}+2e S_z B) \Psi=E^2
\Psi ,
\end{eqnarray}
where $S_z=\frac{i}{2}\gamma_1\gamma_2$ is the spin along $\hat
{\bf z}$.  This equation can also be written in terms of the
orbital angular momentum $z$-component $L_z=-i(\partial_y x-
\partial_x y)$  as
\begin{eqnarray} \label{equMAg3mod} [-{\boldmathnabla}^2
  + \frac{1}{4}e^2 B^2 \rho^2+e {B}(L_z+ 2 S_z )] \Psi=E^2 \Psi ,
\end{eqnarray}
where $\rho^2=x^2+y^2$ is the radial cylindrical coordinate. As
can be seen in the Appendix, the separation of
 this equation  is manifest within these coordinates.
This equation has the  well-known  non-relativistic ($nr$)
Schr\"odinger-equation counterpart of a scalar  particle in a
magnetic field of magnitude $B$\cite{Gasio}
\begin{eqnarray} \label{equMAg3modNR} \frac{1}{2 \mu}(-{\boldmathnabla}^2
  + \frac{1}{4}e^2 B^2 \rho^2+e {B}L_z ) \Psi_{nr}=E_{nr}\Psi_{nr}.
\end{eqnarray}
When $\Psi_{nr}$ is  separated into   cylindrical coordinates
 $\Psi_{nr}= u(\rho)e^{ik_zz}e^{im\phi}$ the radial component $u(\rho)$ satisfies
\begin{eqnarray}
\label{nonrel}
u^{\prime\prime}+\frac{1}{\rho}u^\prime-\frac{m^2}{\rho^2}u-\frac{e^2B^2}{4}\rho^2u+\left
( 2 \mu E_{nr}-k_z^2-eBm\right )u=0
\end{eqnarray} with corresponding energy
\begin{eqnarray}
\label{Landau} E_{nr}=\frac{k_z^2} {2\mu} +\frac{e B}{2\mu}(2
n_r+1+|m|+m) \ \ \ n_r=0,1,...,\ \ \ \
\end{eqnarray}
where  $m$, $n_r$ are quantum numbers related to  $L_z$
 and the radial motion, respectively. The Landau levels
emerge not surprisingly.

 The solution of the  relativistic   Eq. (\ref{equMAg3mod}) follows from
 Eq. (\ref{equMAg3modNR}),
which differs from the former, up to factors, by the spin
operator. Eq.  (\ref{equMAg3mod}) constitutes only a necessary
condition and has in fact more freedom in the  solutions than the
original Eq. (\ref{HamFWeq}) (albeit the energies are the same,
for the latter equation is obtained from the modified Foldy-Wouthuysen unitary
transformation). The eigenfunctions from the original equations
are worked out in the Appendix. The results can be obtained by
using the eigenfunctions of the total angular  momentum component
$L_z+S_z$,  which leads to equations of the form of the massless
($\mu=0$) (A.\ref{decoupledeqs})-(A.\ref{decoupledeqsend})  or
massive
(A.\ref{massdecoupledeqexp})-(A.\ref{massdecoupledeqexpend}). The
energy eigenvalues are
\begin{eqnarray}
\label{massdecoupledeqs} &E=\sqrt{k_z^2+2eB( n_r+ m+1)+\mu^2}\ \ \
& m\geq 0\ \ \ n_r=0,1,...\\ \ \ \ \ \label{massdecoupledeqso}
&E=\sqrt{k_z^2+2 eB n_r+\mu^2}\ \ \ & m< 0\ \ \ n_r=1,2,...\ \ \ \
\end{eqnarray}

\subsection{$5+1\rightarrow 3+1$ reduction }

The procedure we have followed in Eqs.
(\ref{KKmag})-(\ref{equMAg3}) is valid for any dimensional
reduction. In the 5+1 case
\begin{eqnarray}
\label{3dmagend}
 A_1=A_x=0, \ \ \  A_2=A_3=0, \ \ \  A_3=A_z=0, \ \ \
\nonumber \\ \label{A5d} A_4=A_u=-\frac{1}{2} v B,\ \ \
A_5=A_v=\frac{1}{2} u B,
\end{eqnarray}
where we have chosen  5-$d$ coordinate  labels $
(u_1,u_2,u_3,u_4,u_5)=(x,y,z,u,v).$ Generalized terms $\pi_I $
are then obtained from  Eq. (\ref{pi}), using the corresponding
${\mathsf A}_{I}$ terms defined above and  the ${\mathsf E}_I=0$.
%One just needs to check the solution in the actual equation. (*DO IT?)
The Hamiltonian $H^\prime $ (Eq. (\ref{FWtrans}))
 results in
\begin{eqnarray}
\label{KKmag5D} H^\prime= \alpha_0
 {[ (\boldmathalpha\cdot\boldmathpi)^2+(\alpha_I \pi_I)^2+ \mu^2
 ]}^{1/2}  \ \ \ \ \ ( I\ {\rm  summed\  over\  4,5}).
\end{eqnarray}
The $5+1\rightarrow 3+1$ case
 equation,  counterpart to Eq. (\ref{equMAg3}), with the magnetic field in Eq. (\ref{A5d}) has the form
\begin{eqnarray} \label{equMAg3mod6d}
 [-{\bf \boldmathnabla}^2-\partial_u^2-\partial_v^2
  + \frac{1}{4}e^2 B^2 \rho^2+e {B}(L_{45}+ 2 S_{45} )] \Psi=E^2 \Psi ,
\end{eqnarray}
where now  $\rho^2=u^2+v^2$, $L_{45}= -i(\partial_v u-
\partial_u v)$,   $S_{45}=\frac{i}{2}\gamma_4\gamma_5$,
and the  spectrum has a similar form to Eqs.
(\ref{massdecoupledeqs})-(\ref{massdecoupledeqso}), with the
transformation $k_z^2\rightarrow {\bf k}^2$, ${\bf k}$
representing the 3-$d$ momentum, which gives a contribution to
the energy  as a free-particle kinetic term. This is but a
consequence of the {\it decoupling } caused by the fact that the
interaction acts purely on the $u$, $v$ dimensions but leaves the
other free. This in turn results from the form of the interaction
which separates the equations corresponding to Eqs. (\ref{M1}) and
(\ref{M2}).

Thus, we obtain a tower of states, similarly to the Kaluza-Klein
mechanism, but with different associated radii (up to degeneracy),
the minimum being occupied by the ground state. %NEW

The form of Eqs.
(\ref{massdecoupledeqs})-(\ref{massdecoupledeqso})  suggests that
the magnetic field  extra-dimensional parameter may be
interpreted as a mass term in ``real" dimensions. Therefore, we
note that a mass-creation mechanism emerges here, with the masses
appearing with a characteristic spectrum. This mechanism is
possible only in a reduction from even $d$ to $d-2$ dimensions.

Although other Dirac-supersymmetric interactions lead to
compactification, they elude a simple solution treatment as obtained
with the magnetic field. %NEW

\section{Conclusions}

  In this work we have presented a mechanism for compactification
through gauge fields. This mechanism allows for independent
behavior in some dimensions  but forces motion of particles in
the other dimensions to be confined, which amounts to an
effective compactification  (driven by a physical process).
 Simplicity and succinctness are gained for compactification can be ascribed to a
field rather than being assumed. Also, the  familiar gauge fields
can produce this mechanism, without need to invoke others.
Although in this work we have concentrated on the didactic
$3+1\rightarrow 1+1$ and novel $5+1\rightarrow 3+1$ cases, this
mechanism is applicable to any even-dimension reduction
$d+2\rightarrow d$. In addition,  this mechanism is  general in
the sense that it is valid  for fundamental spin-1/2 matter
fields.

The results obtained are general for the compactification
presented  can be relevant both independently of or in relation to
curved space. In the latter case, it is  assumed that these
interactions could be eventually related to the gravitational
field acting in the additional dimensions. %NEW

However, this work remains exploratory for it concentrates more
in showing such a mechanism is possible and one still needs
further generalization to relate it to  a universal interaction.
Fields generated by the hypercharge interaction together with
non-abelian ones as the  electroweak interaction or gravitational
fields are feasible candidates. Some ideas on possible
constraints on allowed dimensions and interactions are found in
Ref. \cite{bespro}. In addition, there is room for additional
types of interaction producing compactificaction. The $U(1)$
magnetic field chosen here preserves translational symmetry which
implies the choice of coordinate around which particles rotate in
the extra dimensions is arbitrary. Further characterization of
the interaction may lead, for example, to a choice of this point
in the compactification plane, which would violate Poincar\'e
invariance in the extra dimensions, without direct influence in
the ``real" ones, just as occurs for branes in string theory.

Another outcome of this work is a possible mass creation
mechanism. We have shown that a mass of a free particle in 4-$d$
space can be generated through an  interaction acting on  5 and 6
dimensions.

The   main lesson from this paper is that it is possible to
construct a compactifying interaction which features dimensional
decoupling, at least for the ``real" dimensions, and which may
have consequences in terms of parameters as the mass, but
otherwise leave the same physics  for free physical particles. It
should be interesting to consider the presence of   interactions
inside ``real" space.

Further work should then deal with non-abelian fields, consider
 other separable interactions,  additional multipoles of the magnetic field, and gravitation,  and try to relate them
to cosmological models, for these fields should appear
self-consistently.

\noindent{\bf Acknowledgments}

The authors
 acknowledge support from  DGAPA-UNAM through project
IN127298. One of us (J. B.) thanks A. de la Macorra for helpful
discussions.

\setcounter{equation}{0}
%\appendix
\noindent{\bf Appendix}

In this Appendix we solve directly Dirac's equation $H\Psi=E\Psi
$ for a particle in a constant magnetic field in the 4-$d$ case
as a complement to Eqs. (\ref{equMAg3mod}) and
(\ref{equMAg3mod6d}), for the massless and massive cases.

{\bf Massless case}

We use the chiral representation for the Dirac matrices. Then,
 the $3+1$ component of the Hamiltonian in Eq. (\ref{HamDS}) leads to
\begin{eqnarray}\label{metric}
 \left( \begin{array}{cc}
\boldmathsigma\cdot(-i{ \boldmathnabla }- {\mathbf A}) & 0  \\
0 & -\boldmathsigma\cdot(-i{  \boldmathnabla}- {\mathbf A})
\end{array} \right)\Psi=E\Psi,
\end{eqnarray}
where, from Eq. (\ref{A3d}),  a constant magnetic along $\hat{\bf
z}$  is given by ${\bf A}=\frac{1}{2}B(-y,x,0).$ We use the
constants of the motion to obtain and classify the solutions.
These comprise a component of the total angular momentum, the
momentum, both  in the $\hat{\bf  z}$ direction, and, of course,
the Hamiltonian. In the massless case the chirality  is also a
constant  of the motion and we use it to   separate  Eq.
(A.\ref{metric}) into its chiral components. From the form of
$\gamma_5$ in the  chiral representation
\begin{eqnarray}\label{gamm5chi}
 \gamma_5=\left( \begin{array}{cc}
I & 0  \\
0 & -I
\end{array} \right)
\end{eqnarray}
the upper and lower parts of $\Psi$
\begin{eqnarray}\label{chiralcom}
 \Psi=\left( \begin{array}{c}
\Psi_1   \\
\Psi_2
\end{array} \right)
\end{eqnarray}
correspond to its positive and negative chirality components,
respectively. We now consider the upper, positive chirality part
while it is clear that the other part is obtained with the
solution interchange $E\rightarrow -E$. Given the constants of
the motion we propose  as Anzats for $\Psi_1$, using cylindrical
coordinates $(\rho,\phi,z)$,  separation into a plane wave along
the $\hat{\bf  z}$ direction with momentum $k_z$, a $J_z$
eigenstate with associated angular momentum $m+1/2$, and radial
functions $f(\rho)$, $g(\rho)$ along the $xy$ plane
\begin{eqnarray}\label{psi1}
 \Psi_1=\left( \begin{array}{c}
f(\rho)e^{ik_zz}e^{im\phi}   \\
i g(\rho)e^{ik_zz}e^{i(m+1)\phi}
\end{array} \right).
\end{eqnarray}
With this form for $\Psi_1$ Eq. (A.\ref{metric}) becomes
\begin{eqnarray}
\label{coupledeqs}
k_z f+g^\prime+\frac{m+1}{\rho}g+e\frac{B}{2}\rho g=E f \\
-k_z g -f^\prime+\frac{m}{\rho}f+e\frac{B}{2}\rho f=E g.
\label{coupledeqsend}
\end{eqnarray}
From Eq. (A.\ref{coupledeqs})  $f$ can be expressed in terms $g$
and its derivative. By substituting this $f$ into Eq.
(A.\ref{coupledeqsend}),  and carrying out a similar  procedure
for $g$ from  (A.\ref{coupledeqsend}), one obtains the decoupled
equations
\begin{eqnarray}
\label{decoupledeqs}
f^{\prime\prime}+\frac{1}{\rho}f^\prime-\frac{m^2}{\rho^2}f-\frac{e^2B^2}{4}\rho^2f+\left ( E^2-k_z^2-eB(m+1) \right )f=0\\
 g^{\prime\prime}+\frac{1}{\rho}g^\prime-\frac{(m+1)^2}{\rho^2}g-\frac{e^2B^2}{4}\rho^2g+\left ( E^2-k_z^2-eBm \right )g=0.\label{decoupledeqsend}
\end{eqnarray}
The solution of these equations is constructed with generalized Laguerre
polynomials of the form $L_{n_r}^{|m|}(x^2)$, $x=\sqrt{eB/2}\
\rho$, and each solution leads to the following eigenfunctions
and energy eigenvalues\cite{Gasio} (see also Eqs.
(\ref{nonrel})-(\ref{Landau})): \noindent For $m\ge 0$
\begin{eqnarray}
\label{gfsol}
f(\rho)&=&x^{|m|}e^{-x^2/2}L_{n_r}^{|m|}(x^2) \nonumber \\
g(\rho)&=&\frac{{\sqrt{2 e B}}}{{E} +
k_z}x^{|m+1|}e^{-x^2/2}L_{n_r}^{|m+1|}(x^2), \ \ \ n_r=0,1,...,\ \
\ \
\end{eqnarray}
with  energy
\begin{eqnarray}
\label{energysol} E=\sqrt{k_z^2+2eB( n_r+  m+1)}.
\end{eqnarray}
The  global coefficients in $\Psi_1$ above and the wave functions
below are arbitrary. To normalize the wave function one uses the
cylindrical radial-component integral
\begin{eqnarray}
\label{inteLaguerre} \int_{0}^{\infty}dy  y^m e^{-y}
L_n^m(y)L_{n^\prime}^m(y)=\left \{ \begin{array}{cc} 0\ \ \ & \ \
\ \ n\neq n^\prime \\ \Gamma(1+m) \left(
\begin{array}{c}
n+m  \\
m
\end{array}  \right )\end{array} \right . &  n= n^\prime .
\end{eqnarray}

For $m< 0$
\begin{eqnarray}
\label{gfsolmm}
f(\rho)&=&x^{|m|}e^{-x^2/2}L_{n_r-1}^{|m|}(x^2)  \nonumber \\
g(\rho)&=& -\frac{{\sqrt{2 e B}}}{{E} + k_z} n_r
x^{|m+1|}e^{-x^2/2}L_{n_r}^{|m+1|}(x^2), \ \ \ n_r=1,2,...,\ \ \ \
\end{eqnarray}
with  energy
\begin{eqnarray}
\label{energysolmm} E=\sqrt{k_z^2+2eB n_r}
\end{eqnarray}
 We see that in the latter case we have an increased
degeneracy on the $m$ values which we  ascribe to the conceling
contributions to the energy of the angular motion and its magnetic
moment opposite to the magnetic field. We note our results differ
from those in the textbook of Ref. \cite{Sakurai}
 which claims the spectrum in Eq. (A.\ref{energysolmm}) (including incorrectly
 the $n_r=0$ state), to be general but which represents only a specific type  of solution.

The negative chirality component $\Psi_2$
\begin{eqnarray}\label{psi2}
 \Psi_2=\left( \begin{array}{c}
h(\rho)e^{ik_zz}e^{im\phi}   \\
i j(\rho)e^{ik_zz}e^{i(m+1)\phi}
\end{array} \right)
\end{eqnarray}
has the  solutions for $m\ge 0$
\begin{eqnarray}
\label{gfsolnegchi}
h(\rho)&=& x^{|m|}e^{-x^2/2}L_{n_r}^{|m|}(x^2),  \nonumber \\
j(\rho)&=& \frac{{\sqrt{2 e B}}}{k_z-{E}
}x^{|m+1|}e^{-x^2/2}L_{n_r}^{|m+1|}(x^2),\ \ \ n_r=0,1,...,\ \ \ \
\end{eqnarray}
with  energy
\begin{eqnarray}
\label{energysolnegchi} E=\sqrt{k_z^2+2eB( n_r+  m+1)}.
\end{eqnarray}
and for $m< 0$
\begin{eqnarray}
\label{gfsolmmnegchi}
h(\rho)&=&x^{|m|}e^{-x^2/2}L_{n_r-1}^{|m|}(x^2) \nonumber \\
j(\rho)&=& -\frac{{\sqrt{2 e B}}}{k_z-{E} } n_r
x^{|m+1|}e^{-x^2/2}L_{n_r}^{|m+1|}(x^2), \ \ \ n_r=1,2,...,\ \ \ \
\end{eqnarray}
with  energy
\begin{eqnarray}
\label{energysolmmnegchi} E=\sqrt{k_z^2+2eB n_r}
\end{eqnarray}

{\bf Massive case}

The massive equation
\begin{eqnarray}\label{metricmass}
\left [ \left( \begin{array}{cc}
\boldmathsigma\cdot(-i{ \boldmathnabla}- {\mathbf A}) & 0  \\
0 & -\boldmathsigma\cdot(-i{\boldmathnabla}- {\mathbf A})
\end{array} \right)+\mu\beta\right ]\Psi=E\Psi.
\end{eqnarray}
 can be solved using the same quantum numbers except that the
 mass term (in the chiral representation)
 \begin{eqnarray}\label{gamm0chi}
 \beta=\gamma_0=\left( \begin{array}{cc}
0& -I  \\
-I & 0
\end{array} \right)
\end{eqnarray}
 mixes the  chirality components included in
$\Psi$: $\Psi_1$ in Eq. (A.\ref{psi1}) and $\Psi_2$ in Eq.
(A.\ref{psi1}), which leads  to the slightly more complicated four
coupled radial equations for $f$,  $g$, $h$,  $j$
%\begin{eqnarray}
\begin{eqnarray}
 \label{masscoupledeqs}
k_z f+g^\prime+\frac{m+1}{\rho}g+e\frac{B}{2}\rho g-\mu h=E f \\
-k_z g -f^\prime+\frac{m}{\rho}f+e\frac{B}{2}\rho f-\mu j=E g \\
\label{masscoupledeqs2} -k_z
h-j^\prime-\frac{m+1}{\rho}j-e\frac{B}{2}\rho j-\mu f=E h \\
\label{masscoupledeqs3} k_z
j+h^\prime-\frac{m}{\rho}h-e\frac{B}{2}\rho h-\mu g=E j.
\label{masscoupledeqsend}
\end{eqnarray}
 These equations decouple into four equations for each function. For example,  by solving for $g$
 in Eq. (A.\ref{masscoupledeqs2}) and for $h$  in Eq. (A.\ref{masscoupledeqs3})  and substituting them
into Eq. (A.\ref{masscoupledeqs}) an  equation only  for $f$ is
obtained; by using similar  procedures for the other functions
one gets for all
\begin{eqnarray}
\label{massdecoupledeqexp}
f^{\prime\prime}+\frac{1}{\rho}f^\prime-\frac{m^2}{\rho^2}f-\frac{e^2B^2}{4}\rho^2f+\left ( E^2-k_z^2-eB(m+1)-\mu^2 \right )f=0\\
 g^{\prime\prime}+\frac{1}{\rho}g^\prime-\frac{(m+1)^2}{\rho^2}g-\frac{e^2B^2}{4}\rho^2g+\left ( E^2-k_z^2-eBm-\mu^2 \right )g=0\\
 h^{\prime\prime}+\frac{1}{\rho}h^\prime-\frac{m^2}{\rho^2}h-\frac{e^2B^2}{4}\rho^2h+\left ( E^2-k_z^2-eB(m+1)-\mu^2 \right )h=0\\
 j^{\prime\prime}+\frac{1}{\rho}j^\prime-\frac{(m+1)^2}{\rho^2}j-\frac{e^2B^2}{4}\rho^2j+\left ( E^2-k_z^2-eBm-\mu^2 \right
 )j=0.\label{massdecoupledeqexpend}
\end{eqnarray}
 The solutions for  $m\ge 0$ are
\begin{eqnarray}
\label{gfsolmass}
f(\rho)&=&x^{|m|}e^{-x^2/2}L_{n_r}^{|m|}(x^2) \nonumber  \\
g(\rho)&=&\frac{{\sqrt{2 e B}}}{{E} +
k_z}x^{|m+1|}e^{-x^2/2}L_{n_r}^{|m+1|}(x^2) \nonumber  \\
h(\rho)&=& -\frac{{\mu}}{{E} +
k_z} x^{|m|}e^{-x^2/2}L_{n_r}^{|m|}(x^2)  \nonumber  \\
j(\rho)&=& 0 , \ \ \ n_r=0,1,...,\ \ \ \
\end{eqnarray}
and for  $m< 0$
\begin{eqnarray}
\label{gfsolmassnegm}
f(\rho)&=&x^{|m|}e^{-x^2/2}L_{n_r-1}^{|m|}(x^2) \nonumber  \\
g(\rho)&=&-\frac{{\sqrt{2 e B}}}{{E} +
k_z} n_r x^{|m+1|}e^{-x^2/2}L_{n_r}^{|m+1|}(x^2)  \nonumber \\
h(\rho)&=& -\frac{{\mu}}{{E} +
k_z} x^{|m|}e^{-x^2/2}L_{n_r-1}^{|m|}(x^2)  \nonumber \\
j(\rho)&=& 0 , \ \ \ n_r=0,1,...,\ \ \ \
\end{eqnarray}
with energies
 as above, except that
these are modified with a mass term, and they are  given in Eqs.
(\ref{massdecoupledeqs}) and (\ref{massdecoupledeqso}).

This 4-$d$ massive case can also serve to solve the 6-$d$
massless equation with a generalized magnetic field. Indeed, we
divide the extended Dirac Hamiltonian equation into the 3-
 and 4- and 5-$d$ components
\begin{eqnarray}
\label{HamDSApe} ( { \boldmathalpha}\cdot{\bf p}+{ \alpha}_I {\bf
\pi}_I)
 \Psi=E\Psi  \ \ \ \ \ ( I\ {\rm  summed\  over\  4,5}),
\end{eqnarray}
%by making the appropriate relations among variables.
where  $\pi_I =p_I-{\mathsf A}_I,$  and the ${\mathsf
 A}_I$  are obtained from Eq. (\ref{3dmagend}).
 The   3-$d$ space components appear with bold type. This
massless equation can be projected into the  two chiral
components by $\frac{1}{2}(1\pm \gamma_7)$, with
 $\gamma_7=-i\alpha_0\alpha_6$ (Eqs. (\ref{recualg}), (\ref{recualgend})) being the matrix
 which anticommutes with all 6-$d$ ${\gamma_\mu} 's.$ Each
projected equation can be written in terms of 4-$d$ matrices. As
these satisfy the same relations as those contained in Eq. (A.\ref
{metricmass}), with the mapping 6-$d\rightarrow 4$-$d$ with
$\frac{1}{2}(1 + \gamma_7){\boldmathalpha} \cdot {\bf
p}\rightarrow \mu \beta$, $\frac{1}{2}(1 + \gamma_7){ \alpha}_I
{\bf \pi}_I \rightarrow { \boldmathalpha}\cdot { \boldmathpi}$,
one obtains the corresponding set of equations as
(A.\ref{masscoupledeqs})-(A.\ref{masscoupledeqs3}) (and similarly
for the other chirality part).

 The equations   solved here can also be useful
to solve the intermediate Eq. (\ref{Eigenfacil}), and reproduce in
fact the separation of variables as described in Section 3.

%¡¡¡ EST\'A MUY ENDOG\'AMICA LA BIBLIOGRAF\'IA !!!

%\hskip

% when applying vacuum non-conserving interactions there is apparently a coupling to higher
%dimensions. Can this be used as a test of higher dimensions?

\end{document}